\documentstyle[12pt]{article}
\begin{document}
\begin{center}
{\Large \bf
Indeterministic Quantum Gravity} \\[0.5cm]
{\large\bf II. Refinements and Developments} \\[1.5cm]
{\bf Vladimir S.~MASHKEVICH}\footnote {E-mail:
mash@phys.unit.no}  \\[1.4cm]
{\it Institute of Physics, National academy
of sciences of Ukraine \\
252028 Kiev, Ukraine} \\[1.4cm]
\vskip 1cm

{\large \bf Abstract}
\end{center}
This paper is a continuation of the paper~\cite{1}.
Indeterministic quantum gravity is a theory that unifies
general relativity and quantum theory involving indeterministic
conception, i.e., quantum jumps. By the same token the theory
claims to describe all the universe.

Spacetime is the direct product of cosmic time and space. The
state of the universe is given by metric, its derivative with
respect to cosmic time, and the number of an energy level. A
quantum jump occurs at the tangency of two levels. Equations
of motion are the restricted Einstein equation (the cosmic
space part thereof) and a probability rule for the quantum jump.

\newpage

\hspace*{6 cm}
\begin{minipage} [b] {9 cm}
We make chance a matter of absolute calculation
\end{minipage}
\begin{flushright}
Edgar Allan Poe \vspace*{0.8 cm}
\end{flushright}

\begin{flushleft}
\hspace*{0.5 cm} {\Large \bf Introduction}
\end{flushleft}

The idea of a possible role of gravity in quantum
indeterminism, i.e., in state vector reduction is not new. It
has been advanced, among others, by Feynman~\cite{2}. A
discussion of this role is contained in the paper and the
book~\cite{3} by Penrose. In the papers~\cite{4} there is a
calculation of a possible coherence breaking due to metric
fluctuations. But it is only in the paper~\cite{1} that the
idea has been embodied in an indeterministic dynamics. To do
this a new physical principle has been advanced --- that of
cosmic energy determinacy.

This paper, being a continuation of the paper~\cite{1},
contains some refinements and develop\-ments. The most
important ones are the following.

Due to quantum jumps, it is only the restricted Einstein
equation that holds, namely, $G^{ij}=(\Psi,T^{ij}\Psi)$,
where $i,j=1=2=3$ correspond to cosmic space.

The state of the universe is $\omega=(g,g',l)$, where $g$ is
metric, $g'$ is its derivative with respect to cosmic time,
and $l$ is the number of an energy level.

A quantum jump occurs at and only at a tangency of two
levels.

Equations of motion reduce to the restricted Einstein equation
and a probability rule for the quantum jump.

\section{The universe as a physical system: Spacetime and
matter}

The universe $U$ as a physical system is a pair:$$U=(st,m),
\eqno{(1.1)}$$where $st$ is spacetime, $m$ is matter. The
spacetime is a triple:$$st=(M,g,\nabla),\eqno{(1.2)}$$where
$M$ is a differentiable 4-dimensional manifold, $g$ is a
metric, and $\nabla$ is the Levi-Civita connection. The
matter is a family $F$ of quantum fields on $M$:$$m=F.
\eqno{(1.3)}$$

\section{The structure of spacetime: Cosmic time and
space}

The manifold $M$ is a direct product of two manifolds:
$$M=T\times S,\quad M\ni p=(t,s),\quad t\in T,\quad s\in S
.\eqno{(2.1)}$$The 1-dimensional manifold $T$ is the cosmic time,
the 3-dimensional one $S$ is the cosmic space. By eq.(2.1) the
tangent space at a point $p\in M$ is$$M_p=T_p\oplus S_p.
\eqno{(2.2)}$$In accordance with this, we have for a tensor
$K=K^{\mu \nu}$ of rank 2 the following decomposition:
$$K=K_T+K_S+K_{TS}+K_{ST},\eqno{(2.3)}$$where$$K_T=K_{TT}=
K^{00},\quad K_S=K_{SS}=K^{ij},\quad K_{TS}=K^{0i},\quad
K_{ST}=K^{i0}.\eqno{(2.4)}$$

For the  metric tensor we have$$g=g_T+g_S,\quad g_{TS}=g_{ST}=0,
\eqno{(2.5)}$$which implies $T\bot S$. Furthermore,
$$g_T=dt\otimes dt,\quad g=dt\otimes dt-\tilde g_t,\eqno{(2.6)}$$
or$$g=dt^2-\tilde g_{t\,ij}dx^idx^j.\eqno{(2.7)}$$

\section{The state of the universe}

The state of the universe is$$\omega=(\omega_{st},\omega_m);
\eqno{(3.1)}$$the state of the spacetime is$$\omega_{st}=
(\tilde g,\tilde g'),\eqno{(3.2)}$$where prime denotes the
derivative with respect to the cosmic time $t$; the state
of the matter is$$\omega_m=(\Psi,\cdot\Psi),\eqno{(3.3)}$$
where $\Psi$ is a state vector. Thus$$\omega=(\tilde g,
\tilde g',(\Psi,\cdot\Psi)).\eqno{(3.4)}$$

\section{Equations of motion}

The equations of motion are$$G_S=\omega_m(T_S)=(\Psi,T_S\Psi),
\quad or\quad G^{ij}=\omega_m(T^{ij}),\eqno{(4.1)}$$
$$H_t\Psi_t=\varepsilon_t\Psi_t,\eqno{(4.2)}$$where $G$ is
the Einstein tensor, $T$ is the Schr\"odinger operator of
the energy-momentum tensor, $H_t$ is the Schr\"odinger
Hamiltonian:$$H_t=\int_S\mu(ds)T_{Ts,t}(\partial/\partial t,
\partial/\partial t),\quad T_T(\partial/\partial t,\partial/
\partial t)=T^{00}.\eqno{(4.3)}$$

Eq. (4.1) is the restricted Einstein equation. The equations
for $T,TS$, and $ST$, i.e., for $0\mu,\mu 0$ components are
not used: The components $G^{0\mu}=G^{\mu 0}$ do not
involve $\tilde g''$, so that the corresponding equations
are violated at quantum jumps, i.e., jumps of $\omega_m$.
This violation is a manifestation of quantum holism
connected with the character of quantum state.

To ensure against misunderstanding, it should be stressed
that the restriction of the Einstein equation does not
violate the covariance (or, more precisely, the geometric
character) of the theory developed --- in view of the
structure of the spacetime, eq. (2.1).

Eq. (4.2) realizes the cosmic energy determinacy principle.

There are 6+1=7 equations (4.1),(4.2) for 6+1=7 quantities
$\tilde g_{ij},\Psi$.

\section{The equation for state vector between jumps}

Let us consider eq.(4.2) in more detail. We have for the
Hamiltonian$$H_t=\sum_{l}\varepsilon_{lt}P_{lt},
\eqno{(5.1)}$$$P_{lt}$ being the projector for a level $l$.
Then eq.(4.2) boils down to$$P_{lt}\Psi_t=\Psi_t.
\eqno{(5.2)}$$ Designating $P_l=P$, we have between
jumps$$\Psi=P\Psi,\quad \Psi+d\Psi=(P+dP)(\Psi+d\Psi),
\eqno{(5.3)}$$whence$$(I-P)d\Psi=(dP)\Psi.\eqno{(5.4)}$$
This leads to$$P(dP)\Psi=0.\eqno{(5.5)}$$From the condition
$\parallel\Psi+d\Psi\parallel=1$ it follows$$Re\{(\Psi,
Pd\Psi)\}=0.\eqno{(5.6)}$$To determine $Pd\Psi$ we put
$$Pd\Psi=\alpha(dP)\Psi\eqno{(5.7)}$$so that $dP=0$ would
imply $d\Psi=0$. By eqs.(5.7),(5.5)$$Pd\Psi=0.\eqno{(5.8)}$$
Thus$$d\Psi=(dP)\Psi,\quad \frac{d\Psi}{dt}=\frac{dP}{dt}
\Psi.\eqno{(5.9)}$$

The solution to eq.(5.9) is$$\Psi_{t_2}=U(t_2,t_1)\Psi_{t_1},
\quad U(t_2,t_1)=Te^{\int_{t_1}^{t_2}dP_t},\eqno{(5.10)}$$
relations$$U^+(t_2,t_1)=U^{-1}(t_2,t_1)=U(t_1,t_2)=T^+
e^{\int_{t_1}^{t_2}dP_t}.\eqno{(5.11)}$$being fulfilled.

\section{Watched-pot dynamics}

It is interesting to note that the result (5.9) may also be
obtained in the language of quantum jumps. We have an
orthogonal decomposition for $\Psi=P\Psi$:$$\Psi=(P+dP)\Psi+
(I-P-dP)\Psi=(P+dP)\Psi+(-dP)\Psi.\eqno{(6.1)}$$For the
corresponding probabilities in the first order, relations
$$\parallel(-dP)\Psi\parallel^2=0,\quad \parallel(P+dP)\Psi
\parallel^2=1\eqno{(6.2)}$$hold, so that$$\Psi\to(P+dP)\Psi,
\eqno{(6.3)}$$whence eq.(5.9) results.

Such a dynamics is analogous to the watched-pot one,
the role of a watching cook being played by the principle
of cosmic energy determinacy.

\section{Absence of degeneracy}

For a general metric, the probability for a level $l$ to be
degenerate is zero, so that degeneracy occurs only at level
crossings. Therefore we have$$TrP_l=1,\quad \omega_m=(\Psi_l,
\cdot\Psi_l)=Tr\{P_l\quad \cdot\quad\},\eqno{(7.1)}$$and
$\omega_m$ reduces to $l$. Thus a state of the universe is
$$\omega=(\tilde g,\tilde g',l).\eqno{(7.2)}$$

\section {Quantum jump}

Let us consider two crossing levels, $l=1,2$. The part of
the Hamiltonian for these is$$\tilde H_t=\varepsilon_{1t}
P_{1t}+\varepsilon_{2t}P_{2t}.\eqno{(8.1)}$$Put crossing
time $t_{cross}=0$, then$$\varepsilon_{10}=\varepsilon_{20}
=\varepsilon_0.\eqno{(8.2)}$$We write $\tilde H_t$ in the
form of$$\tilde H=\varepsilon_2P+\nu P_1,\eqno{(8.3)}$$
where$$P=P_1+P_2,\quad \nu=\varepsilon_1-\varepsilon_2.
\eqno{(8.4)}$$From eq.(8.3) it follows$$\tilde H'=
\varepsilon'_2P+\varepsilon_2P'+\nu'P_1+\nu P'_1.\eqno
{(8.5)}$$Let us introduce the designation$$\Delta f=f_{+0}
-f_{-0}.\eqno{(8.6)}$$We put$$f_0=f_{-0},\eqno{(8.7)}$$
so that$$\Delta f=f_{+0}-f_0.\eqno{(8.8)}$$Relations
$$\Delta\tilde g=0,\quad \Delta g'=0\eqno{(8.9)}$$hold; as
$$H=H[\tilde g,\tilde g'],\eqno{(8.10)}$$we have$$\Delta H
=0.\eqno{(8.11)}$$Indeed,$$\Delta\tilde H=\varepsilon_{2+0}
P_{+0}-\varepsilon_0P_0+\nu_{+0}P_{1+0}-\nu_0P_{10},\eqno
{(8.12)}$$and$$\varepsilon_{2+0}=\varepsilon_0,\quad P_{+0}
=P_0,\quad\nu_{+0}=\nu_0=0,\eqno{(8.13)}$$which leads to
eq.(8.11).

As$$H'=H'[\tilde g,\tilde g',\tilde g''],\eqno{(8.14)}$$if
$\Delta\omega_m=0$ then $\Delta\tilde g''=0$ and$$\Delta H'
=0\eqno{(8.15)}$$must hold. We find$$\Delta H'=\Delta
(\varepsilon'_2P)+\Delta(\varepsilon_2P')+\Delta(\nu'P_1)+
\Delta(\nu P'_1),\eqno{(8.16)}$$$$\Delta(\varepsilon_2P')=0,
\quad \Delta(\varepsilon'_2P)=(\Delta\varepsilon'_2)P_0.
\eqno{(8.17)}$$As $\lim_{t\to 0}\nu_t=0$, we put $\Delta(\nu
P'_1)=0$. Then$$\Delta\tilde H'=(\Delta\varepsilon'_2)P_0+
\nu'_{+0}P_{1+0}-\nu'_0P_{10}=0.\eqno{(8.18)}$$Hence we
obtain$$(\Delta\varepsilon'_2-\nu'_0)P_{10}=-\nu'_{+0}P_
{1+0}P_{10},$$$$(\Delta\varepsilon'_2)P_{20}=-\nu'_{+0}
P_{1+0}P_{20}.\eqno{(8.19)}$$There are three solutions to
eqs.(8.19):$$Ia.\quad P_{1+0}P_{20}=0,\,P_{1+0}P_{10}=P_{10},
\quad P_{1+0}=P_{10},\,P_{2+0}=P_{20},$$$$\Delta\varepsilon'_2
=0,\,\nu'_{+0}=\nu'_0,\quad\varepsilon'_{2+0}=\varepsilon'_{20},
\,\varepsilon'_{1+0}=\varepsilon'_{10}.\eqno{(8.20)}$$$$Ib.
\quad P_{1+0}P_{20}=P_{20},\,P_{1+0}P_{10}=0,\quad P_{1+0}=
P_{20},\,P_{2+0}=P_{10},$$$$\Delta\varepsilon'_2=-\nu'_{+0},\,
\Delta\varepsilon'_2=\nu'_0,\quad\varepsilon'_{2+0}=
\varepsilon'_{10},\,\varepsilon'_{1+0}=\varepsilon'_{20}.
\eqno{(8.21)}$$$$II.\quad P_{1+0}P_{20}\ne 0,P_{20},\quad
P_{1+0}P_{10}\ne 0,P_{10},\quad\{P_{1+0},P_{2+0}\}\ne\{P_{10},
P_{20}\},$$$$\Delta\varepsilon'_2=0,\,\Delta\varepsilon'_2-
\nu'_0=0,\,\nu'_{+0}=0,\quad\varepsilon'_{2+0}=\varepsilon'_
{1+0}=\varepsilon'_{20}=\varepsilon'_{10}.\eqno{(8.22)}$$

The solutions $Ia$ and $Ib$ in fact coincide and describe a
simple crossing, i.e., that without tangency. In that case
there is no jump.

The solution $II$ describes a crossing with tangency. In case
of such a crossing, the probability for $\{P_{1+0},P_{2+0}\}
=\{P_{10},P_{20}\}$ is equal to 0, so that if there is a
tangency, a jump occurs with the probability 1.

Thus a criterion for the quantum jump is the level tangency.

We have$$\omega_{t_{tang}}=(\tilde g_{t_{tang}},\tilde g'
_{t_{tang}},l_{t_{tang}})\to\omega_{t_{tang}+0}=(\tilde g_
{t_{tang}},\tilde g'_{t_{tang}},l^j_{t_{tang}+0}),$$
$$W_{l^jl}=Tr\{P_{l^j_{t_{tang}+0}}P_{l_{t_{tang}}}\},\quad
l=1,2,\quad l^j=1,2,\eqno{(8.23)}$$where $W_{l^jl}$ is the
probability for the jump $l\to l^j$.

\section{Indeterministic dynamics}

The indeterministic dynamics constructed above is very simple
in the conceptual sense (but only in that sense!): Its only
ingredients are the state of the universe$$\omega=(\tilde g,
\tilde g',l),\eqno{(9.1)}$$the restricted Einstein equation
$$G_S=Tr\{P_lT_S\}\eqno{(9.2)}$$for time evolution between
jumps, and jumps at the level tangency$$(\tilde g,\tilde g',
l)\to (\tilde g,\tilde g',l^j),\quad W_{l^jl}=Tr\{P_{l^j}
P_l\}.\eqno{(9.3)}$$The dynamics is unified: Both
deterministic part and quantum jumps follow from the principle
of cosmic energy determinacy. Further still, strange though it
might seem, all the dynamics may be described in terms of
quantum jumps.

Indeterministic dynamics involves some subtleties, therefore
there is good reason to describe it in more detail.

For $t_1,t_2$ between two jumps, $t_2>t_1$, we introduce a
designation for deterministic time evolution:
$$\omega_{t_2}=D(t_2,t_1)\omega_{t_1}.\eqno{(9.4)}$$
In addition, we introduce an auxiliary deterministic time
evolution with a fixed initial state of the matter:
$$\omega_{t_2t_1}=D_{t_1}(t_2,t_1)\omega_{t_1},\quad\omega
_{mt}=\omega_{mt_1}\quad for\quad t_1\le t\le t_2.\eqno
{(9.5)}$$

For $t=t_{cross}=0$ we have
$$\omega_0=\{\tilde g_0,\tilde g'_0,P_{l0}\},\quad l=1\;
or\;2.\eqno{(9.6)}$$
Then for $\delta t>0$
$$\omega_{\delta t\,0}=D_0(\delta t,0)\omega_0=\{\tilde g_{
\delta t\,0},\tilde g'_{\delta t\,0},P_{l0}\}.\eqno{(9.7)}$$
For this state, the Hamiltonian $\tilde H$ is
$$\tilde H_{\delta t\,0}=\varepsilon_{1\,\delta t\,0}P_{1\,
\delta t\,0}+\varepsilon_{2\,\delta t\,0}P_{2\,\delta t\,0},
\eqno{(9.8)}$$and we put for $t=\delta t+0$
$$\omega_{\delta t+0\,0}=\{\tilde g_{\delta t\,0},\tilde g'_
{\delta t\,0},P_{l^j\,\delta t\,0}\},\quad l^j=1\;or\;2.\eqno
{(9.9)}$$
For $t=+0$, a true state is by definition
$$\omega_{+0}=\omega_{+0\,0}=\lim_{\delta t\to+0}\omega_{\delta
 t+0\,0}=\{\tilde g_0,\tilde g'_0,P_{l^j\,+0}\},\quad P_{l^j\,
+0}=\lim_{\delta t\to+0}P_{l^j\,\delta t\,0}.\eqno{(9.10)}$$
This gives the explicit definition for $P_{l^j\,+0}$.

For $t>0$, the true state is given by
$$\omega_t=\lim_{\delta t\to+0}D(t,\delta t)\omega_{\delta
t+0\,0}.\eqno{(9.11)}$$
It is important to stress that
$$\omega_t\ne D(t,+0)\omega_{+0}.\eqno{(9.12)}$$
The value
$$\omega_{+0}=\lim_{t\to+0}\omega_t\eqno{(9.13)}$$
coincides with that given by eq.(9.10).

\section{Deterministic approximation}

The relations$$divG=G^{ij}{}_{;i}=0,\eqno{(10.1)}$$$$divT_H=0
\eqno{(10.2)}$$hold ($H$ denotes the Heisenberg picture).

In the deterministic approximation, i.e., with no quantum
jumps we have$$\Psi_H=const.\eqno{(10.3)}$$Then$$div(\Psi,
T\Psi)=div(\Psi_H,T_H\Psi_H)=(\Psi_H,divT_H\Psi_H)=0,\eqno
{(10.4)}$$and the (unrestricted) Einstein equation$$G=(\Psi,
T\Psi)\eqno{(10.5)}$$may be fulfilled with$$G^{0\mu}=(
\Psi,T^{0\mu}\Psi)\eqno{(10.6)}$$as initial conditions.

Thus in the deterministic approximation, quantum holism
does not manifest itself.

\end{document}